\documentclass[twocolumn,twoside,slac_two]{revtex4}
\usepackage{graphicx}
\usepackage{fancyhdr}
\usepackage{amssymb}
\begin{document}

\title{Status of MAGIC-II}

%

\author{A. Moralejo}
\affiliation{Institut de F{\'{\i}}sica d'Altes Energies, Bellaterra,
  Barcelona 08193, Spain}
\author{for the MAGIC collaboration}

\begin{abstract}
A status report of the second phase of the MAGIC ground-based
gamma-ray facility (as of October 2009) is presented. MAGIC became
recently a stereoscopic Cherenkov observatory with the inauguration of
its second telescope, MAGIC-II, which is currently approaching the end
of its commissioning stage.

\end{abstract}

\maketitle

\thispagestyle{fancy}


\section{INTRODUCTION}
MAGIC, a gamma-ray imaging atmospheric Cherenkov facility \cite{magic}
composed up to now of a single telescope, has become a stereoscopic
system with 
the inclusion of a second telescope, MAGIC-II. MAGIC is located on the
Canary island of La Palma (28.8$^\circ$ N, 17.9$^\circ$ W), and
operates in the very high energy spectral band (photon energies above
30 GeV). MAGIC-II saw its first light in Spring 2009, and is now
approaching the end of its commissioning phase. Whereas from the
mechanical point of view MAGIC-II is essentially a clone of the first
telescope, it features significant improvements in other aspects, like
a more finely pixelized camera and a lower-cost, more compact readout
system.

\section{THE MIRROR DISH}
\begin{figure*}[ht]
\centering
\includegraphics[width=0.85\textwidth]{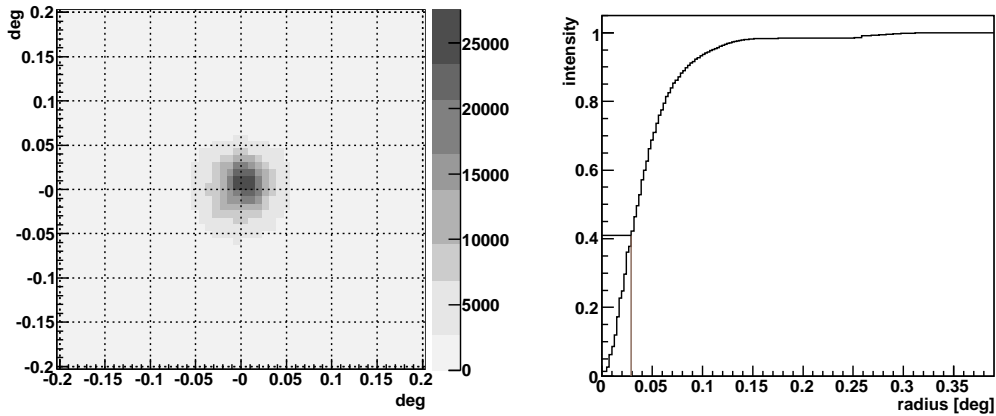}
\caption{Optical point spread function of MAGIC-II. The radius of the
  camera pixels is 0.05$^\circ$. 66$\%$ of the light from a point
  source impinging paralel to the optical axis of the telescope is
  collected within one pixel.} \label{M2PSF}
\end{figure*}

The mirror dish of MAGIC-II is a tessellated paraboloid of 17 m focal
length and f/D=1. Each of the 247 square tiles is a spherical mirror
with a surface of 1 m$^2$, mounted on two motors which allow to adjust
its orientation to ensure that the parabolic shape is maintained,
despite the sagging of the telescope structure, for different
orientations of the instrument. The reflecting surfaces of the inner
143 mirror tiles \cite{mirrors1} are diamond-milled aluminum plates
(the same technology used in MAGIC-I), whereas the outer 104 are
equipped with thin aluminum-coated glass sheets \cite{mirrors2}; in
both types of mirrors the needed mechanical stiffness is provided by
an underlying aluminum honeycomb structure. The optical properties of
MAGIC-II are similar to those of the first telescope. About 66$\%$ of
the light from a point source is contained within the area of one
camera pixel (see fig. \ref{M2PSF}). The good optical quality of the
mirror dish is confirmed by the analysis of muon ring images (a
typical one is shown in fig. \ref{muonring2}).

\section{CAMERA AND DATA ACQUISITION}
\begin{figure*}[ht]
\centering
\includegraphics[width=0.8\textwidth]{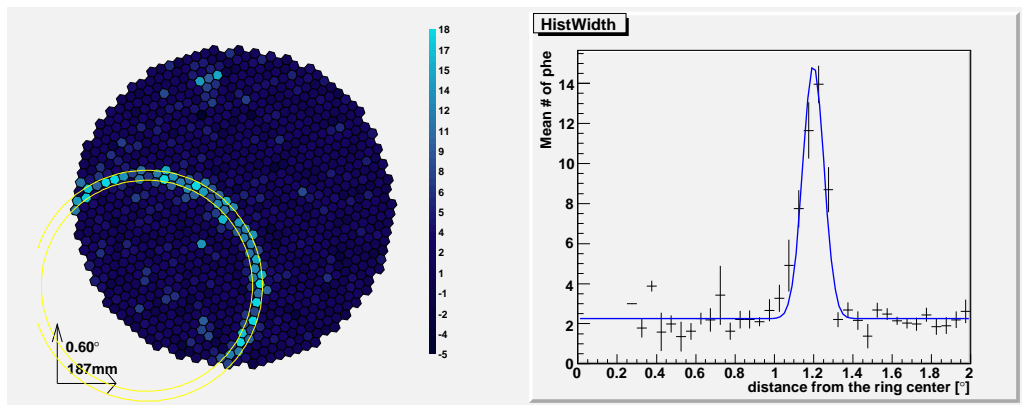}
\caption{A muon ring recorded in September 2009. On the right pad, the
  transversal light profile of the ring. The width of the ring
  confirms the good optical properties of the MAGIC-II mirror dish.} \label{muonring2}
\end{figure*}

The camera  of MAGIC-II (see left pad of fig. \ref{muonring2}) is
composed of 1039 photomultiplier tubes of 0.1$^{\circ} \varnothing$,
for a total field of view of 3.5$^\circ$. The innermost 559 pixels
take part in the trigger (covering a 70$\%$ larger area than in M-I).
\par
The signals from the 1039 PMTs on the MAGIC-II
camera are converted into analog optical pulses which are sent to the
control building 80 m away via optical fibres, where they are
converted back into electronic pulses and then split to provide the
input for the trigger and the signal sampling systems. The readout
system of MAGIC-II is based on version 2 of the Domino Ring Sampler
chip (DRS2). The chip samples the input signals analogically at 2
Gsample/s using an array of 1024 capacitors (so-called cells). In case
of a trigger, the sampling is stopped and the data are digitized with
a 12-bit resolution ADC at 40 MHz. Data management is based on a
digital board called PULSAR that handles up to 80 analog channels. 80
samples (one every 0.5 ns) are recorded per pixel for each triggered
event.
\cite{readout}.
\par
\begin{figure}[ht]
\begin{center}
\includegraphics[width=0.9\columnwidth]{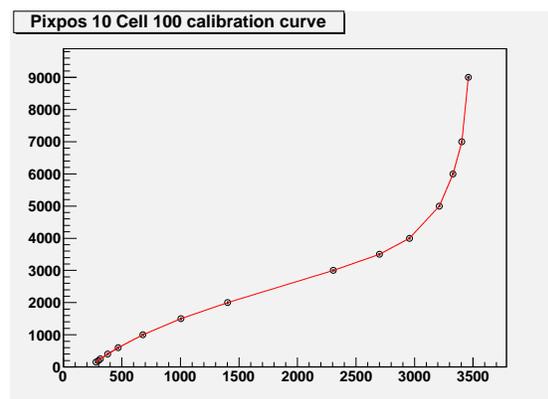}
\caption{Typical linearity calibration curve of a cell in a DRS2 channel. The input
  signal (in units of 0.1 mV) is in the vertical axis, while the
  horizontal one shows the output value in ADC
  counts.} \label{oldcalcurve}
\end{center}
\end{figure}

The response of DRS2 chip is not linear, and amplitudes have to be
corrected as the first step in the processing of the data. A dedicated
calibration of every capacitor in every channel ($>10^6$ instances ) is
performed once per night for this purpose (see
fig. \ref{oldcalcurve}). Saturation of the signal occurs at about 900
photoelectrons. 

\par
Occasionally, data are contaminated by sharp regular spikes (in
isolated cells) which can be identified and have to be cured by
software, through interpolation of the values in neighboring cells,
as part of the signal extraction process. This effect is rare,
affecting just about 3$\%$ of the measurements, and does not spoil the
data in a significant manner.
\par
\begin{figure}[t]
\begin{center}
\includegraphics[width=0.9\columnwidth]{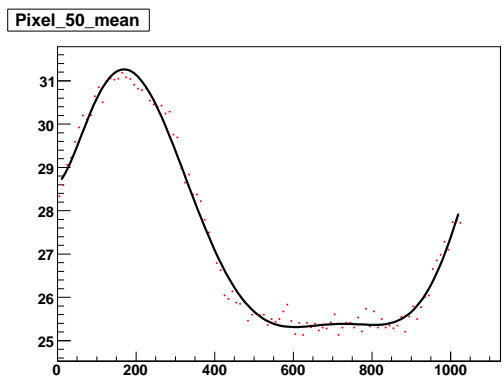}
\caption{Dependence of the arrival time (y-axis, units: 0.5 ns) of
  laser calibration pulses in a certain channel with the number of the
  first DRS2 cell written out (shown in the x-axis). This behaviour is
  characteristic of the DRS2 chip, and can be fully corrected offline
  in order to get the right pulse arrival times in every
  pixel.} \label{pix50_corr}
\end{center}
\end{figure}

A cell-dependent correction of the signal timing is also needed. In
figure \ref{pix50_corr} the average reconstructed arrival time of laser
calibration pulses (fast flashes which illuminate uniformly the whole
camera) is shown for a given channel as a function of the number of
the first DRS 
cell written out in each event. The observed modulation is due to the
non-uniform speed of the ``Domino wave'' that determines which cell
is sampling the input signal at a given time. The curve shown in
fig. \ref{pix50_corr} is characteristic of each channel and stable in
time. One such curve per channel is used for the offline correction of
the signal arrival times.

\section{FIRST OBSERVATIONS}

Stereoscopic observations of the Crab Nebula are being performed with
the two MAGIC telescopes since September 2009. The stereo trigger
system (with orientation-dependent adjustable signal delays) is in
operation since November, but the results shown below correspond to
earlier observations, carried out in ``software stereo'' mode (with
each telescope recording events independently, followed by offline
matching of the events).

\subsection{Data analysis}
After a conventional two-level image cleaning procedure, a simple
geometrical reconstruction has been applied to obtain an estimate of
the shower axis geometry (direction and impact point) as the
intersection of two planes $-$ one per telescope\footnote{More 
sophisticated methods, involving the use of the pixel timing
information, which proved useful in the analysis of single telescope
data \cite{timing}, are being investigated, and preliminary 
results are promising.}. The background suppression relies in the
{\it Random Forest} algorithm \cite{randomforest}, fed with image
parameters from the individual telescopes and also with shower
parameters obtained from the stereoscopic reconstruction (like the
shower impact point and the height of the shower maximum). The {\it
  Random Forest} is trained on a sample of Monte Carlo gammas and real
background events (from observations of an empty sky region), and then
applied to the Crab Nebula data in order to obtain for each event a
single value, dubbed {\it hadronness}, which will be used as cut
parameter for background rejection.

\subsection {Preliminary results}
\begin{figure}[ht]
\begin{center}
\includegraphics[width=0.9\columnwidth]{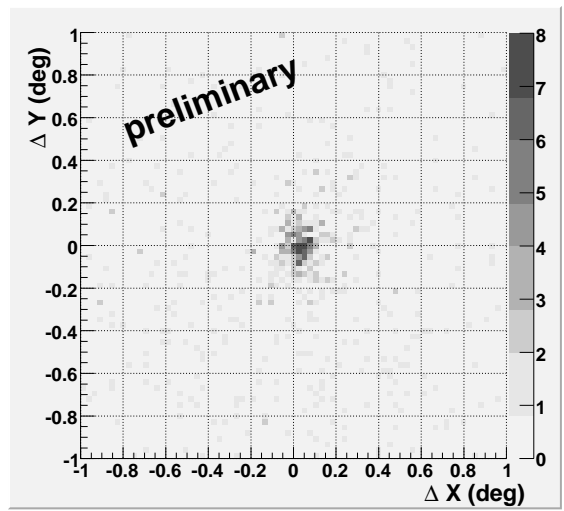}
\caption{Distribution of reconstructed event directions in camera
  coordinates, with (0, 0) corresponding to the position of the Crab
  Nebula. The event sample is the same of
  fig. \ref{Theta2Gt400}} \label{MapSizeGt400}
\end{center}
\end{figure}

\begin{figure}[b]
\begin{center}
\includegraphics[width=0.9\columnwidth]{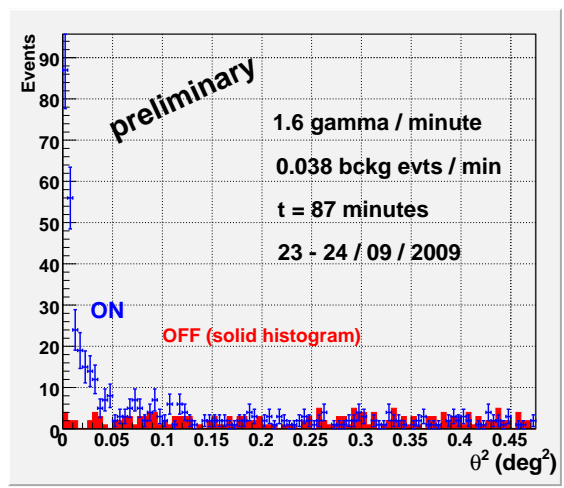}
\caption{Angular distribution of excess events around the Crab Nebula
  ($\theta^2$ is the squared angular distance between the
  reconstructed event direction and the nominal source direction). The
  event sample here displayed (see text) corresponds to a peak gamma
  energy of around 400 GeV.} \label{Theta2Gt400}
\end{center}
\end{figure}

A map of reconstructed event directions (in camera coordinates) with
respect to the Crab Nebula is shown in fig. \ref{MapSizeGt400} for 87
minutes of {\it wobble} observations (i.e. with the telescope pointing
0.4$^\circ$ away from the source direction). The corresponding angular
distribution of events around the source is presented in
fig. \ref{Theta2Gt400}. No offline pointing correction has been
applied. The selected event sample includes only events with at least
400 photoelectrons in each of the telescopes, resulting in a peak
gamma energy $\simeq$ 400 GeV. It is therefore a high energy sample,
very much above the energy threshold of the instrument (of around 50
GeV), but roughly where the best integral flux sensitivity is expected.
\par
In terms of flux sensitivity, the observed performance of the
stereoscopic system is already clearly superior to that of MAGIC-I in
standalone observations, and is approaching the Monte Carlo
expectations. Precise tuning of the Monte Carlo simulation to the
characteristics of MAGIC-II is ongoing, and is expected to improve
further the performance of the system. 

\section {CONCLUSIONS}

The second telescope of the MAGIC ground-based gamma-ray observatory
is already operational, and is performing stereoscopic
observations since September. Preliminary results of the observations
of the Crab Nebula show already a very significant boost in
performance with respect to the standalone MAGIC-I telescope,
particularly in terms of flux sensitivity. Work is ongoing in
polishing the analysis methods and improving the agreement of the
Monte Carlo simulation with data, both of them aspects in which there
is still some room for improvement. 

\bigskip 

\end{document}